\documentclass[a4paper]{article}

\usepackage{INTERSPEECH2022}
\usepackage{multirow}
\usepackage{graphicx}
\usepackage{longtable}
\usepackage{caption}
\usepackage{subcaption}

\title{Anti-Spoofing Using Transfer Learning with Variational Information Bottleneck}
\name{Youngsik Eom, Yeonghyeon Lee, Ji Sub Um, Hoirin Kim}
\address{
  School of Electrical Engineering, KAIST, Daejeon, Republic of Korea
  }
\email{\{iron5605,glory20h,twiz0311,hoirkim\}@kaist.ac.kr}

\begin{document}

\maketitle

\begin{abstract}
  Recent advances in sophisticated synthetic speech generated from text-to-speech (TTS) or voice conversion (VC) systems cause threats to the existing automatic speaker verification (ASV) systems. Since such synthetic speech is generated from diverse algorithms, generalization ability with using limited training data is indispensable for a robust anti-spoofing system. In this work, we propose a transfer learning scheme based on the wav2vec 2.0 pretrained model with variational information bottleneck (VIB) for speech anti-spoofing task. Evaluation on the ASVspoof 2019 logical access (LA) database shows that our method improves the performance of distinguishing unseen spoofed and genuine speech, outperforming current state-of-the-art anti-spoofing systems. Furthermore, we show that the proposed system improves performance in low-resource and cross-dataset settings of anti-spoofing task significantly, demonstrating that our system is also robust in terms of data size and data distribution.
\end{abstract}

\section{Introduction}

Automatic speaker verification (ASV) system verifies whether the claimed speaker is permitted in the system using an utterance from the speaker. Recent advances in speech generation tasks such as text-to-speech (TTS) and voice conversion (VC) pose threats to the security of the ASV system. The anti-spoofing system is designed to solve the security problem of the ASV system and to determine whether the speech is genuine or spoofed at the front end of the ASV system. Training an anti-spoofing system requires a variety of synthetic speech generated from diverse algorithms. However, it is not realistic to construct a training dataset considering all possible attack scenarios because it is not known which TTS or VC system the attacker will use to generate spoofed speech. For this reason, generalization ability using limited training data is important to detect various spoofing attacks from unseen systems in anti-spoofing tasks.

Recent anti-spoofing systems attempt to improve performance on unseen spoofing attacks by constructing elaborate network architectures and features. In particular, many studies have been conducted for effective network architectures while using raw waveform as input to get better representation \cite{ma21d_interspeech,tak21_asvspoof,hua2021towards}. However, the problem of limited training data still causes limitations to generalization ability of the representation that the system can learn. To address this, some recent systems focus on finding the general representation through transfer learning using a pretrained model at the front end of the existing system \cite{wang2021investigating,tak2022automatic}.

Transfer learning is a learning method in which the knowledge of the model is transferred to the target domain by training a huge model with a large amount of data and fine-tuning it with the data of the target task. Xie \textit{et al}. \cite{xie21_interspeech} use transfer learning with pretrained wav2vec \cite{schneider19_interspeech} model and adopt siamese network in anti-spoofing task. However, a large-scale pretrained model can be easily overfitted to the dataset through transfer learning and may learn superficial features which are not intended for the task rather than generalized features \cite{mahabadi2021variational}. Mahabadi \textit{et al}. \cite{mahabadi2021variational} propose to use the information bottleneck (IB) \cite{tishby99information} principle for transfer learning to address the overfitting problem of the pretrained language model. Inspired by \cite{mahabadi2021variational}, we incorporate the IB principle with transfer learning using a pretrained model. IB starts from the idea of finding relevant information from the original signal. Finding a shortcode from the signal that conserves information about the target task label while compressing the original signal provides suppression of irrelevant features with the target task. We adopt variational information bottleneck (VIB) \cite{Alemi17} to implement IB principle. VIB provides a regularization term to the loss function that suppresses irrelevant information from latent representation. Adding this VIB module after the pretrained model helps to extract generalized representations by compressing only meaningful information related to the task.

In this paper, we aim to improve generalization performance for speech anti-spoofing task by using a transfer learning scheme based on the wav2vec 2.0 \cite{baevski2020wav2vec} pretrained model with variational information bottleneck (VIB). The pretrained wav2vec 2.0 model extracts speech embedding based on the representations which are already learned from another task. Then, we exploit the VIB structure which maps the speech embedding to latent feature $z$, regularizing latent feature to learn generalized information by suppressing superficial and redundant information. The regularization term for VIB is added to the training loss during the training phase. Evaluation on ASVspoof 2019 logical access (LA) database \cite{wang2020asvspoof} shows that our proposed system improves performance in terms of minimum normalized tandem detection cost function (min t-DCF) \cite{kinnunen18b_odyssey} and equal error rate (EER), surpassing the performance of prior works. And we show that our system helps to improve performance in low-resource and cross-dataset settings, demonstrating that our system is robust in terms of training data size and data distribution.

\section{Method}

\begin{figure*}[t]
  \centering
  \includegraphics[width=\linewidth,height=5cm]{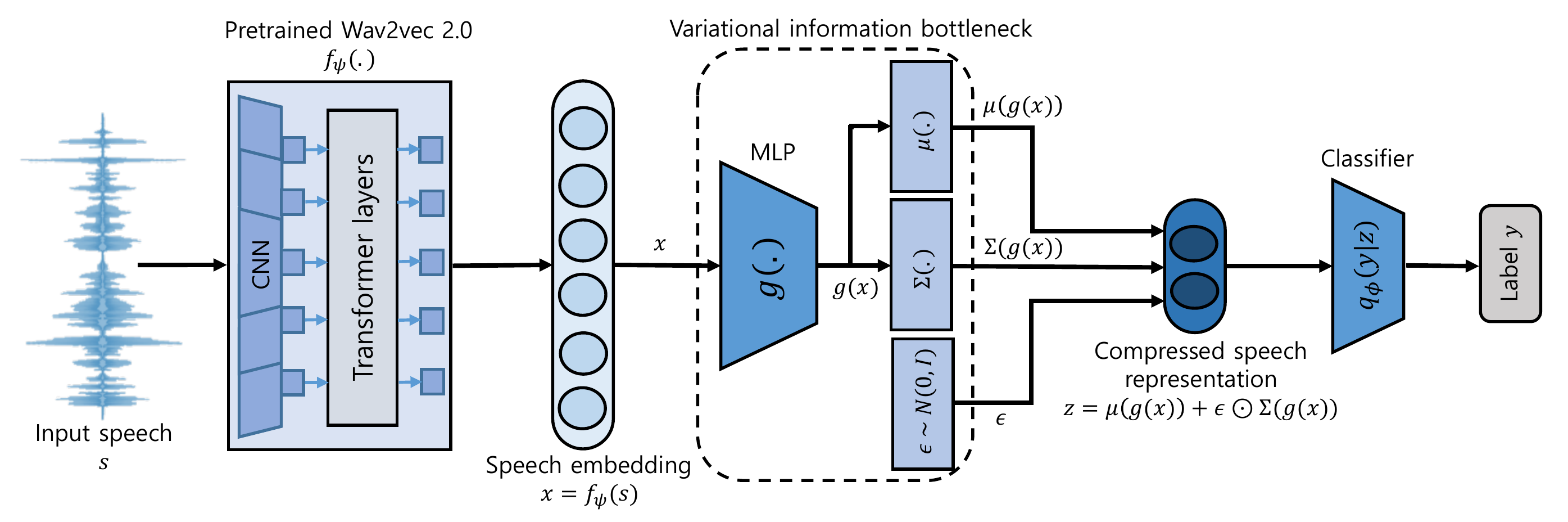}
  \caption{Overall architecture of proposed Wav2vec 2.0 + VIB system.}
  \label{fig:wav2+VIB}
\end{figure*}

\subsection{Wav2vec 2.0}
Wav2vec 2.0 \cite{baevski2020wav2vec} is a framework of self-supervised learning which had recently achieved prominent results in speech recognition. Overall architecture consists of a convolutional neural network (CNN) feature extractor, a context network, and a quantization module. The CNN feature extractor takes raw audio as an input and converts it into a sequence of latent feature vectors. The context network is a series of transformer blocks \cite{vaswani2017attention}. It takes the latent feature vectors generated by the CNN feature extractor and maps them into contextualized representations of speech. The quantization module of the wav2vec 2.0 model is used to generate quantized latent representations for the pretraining objective.

The framework consists of two phases. In the first phase, the model is pretrained with large amounts of unlabeled speech data using a contrastive objective, which is similar to a masked-language modeling objective in BERT \cite{devlin2018bert}. As a result of pretraining, the model learns to emit speech embeddings that contain contextualized information for a given input speech. In the second phase, a prediction head with a simple structure is added to the output of the pretrained wav2vec 2.0 model, then the whole model with the prediction head is fine-tuned for various downstream tasks such as speech recognition. The fine-tuning is conducted using relatively small amounts of task-specific labeled data. This scheme of pretraining and fine-tuning makes it possible to leverage the knowledge that can be learned from large amounts of unlabeled speech, which are generally easier to collect compared to labeled speech. Although the original work was proposed for speech recognition, many works have explored the use of wav2vec 2.0 for other speech-related tasks and the framework has proven that it shows outperforming performance even when used for other speech tasks \cite{fan21_interspeech, pepino21_interspeech}. For this reason, we adopt wav2vec 2.0 model to leverage pretrained knowledge to the anti-spoofing task.

\subsection{Variational information bottleneck}
According to the IB principle, finding the most compressed latent representation $Z$ from input $X$ and simultaneously preserving information on label $Y$ is done by minimizing the following objective function:
\begin{equation}
  \mathcal{L}_{IB} = \beta I(X,Z) - I(Z,Y),
  \label{eq1}
\end{equation}
where $I(X,Z)$ means the mutual information between input and latent representation, and $I(Z,Y)$ means the mutual information between latent representation and output. $\beta$ is a hyper-parameter for adjusting the ratio between $I(X,Z)$ and $I(Z,Y)$.

Utilizing Equation (\ref{eq1}) in the training scheme directly is difficult since computing mutual information is challenging. Alemi \textit{et al}. \cite{Alemi17} present a variational approximation of IB objective, which is called variational information bottleneck (VIB) to optimize the objective function directly using stochastic gradient descent. The objective function of VIB is derived from the upper bound of the IB objective. Since $\mathcal{L}_{IB} \leq \mathcal{L}_{VIB}$, we can optimize Equation (\ref{eq1}) indirectly by minimizing the following objective function:
\begin{equation}
\begin{aligned}\label{eq2}
\mathcal{L}_{VIB} = \beta \mathbb{E}_{x}&[KL(p_{\theta}(z|x),r(z))] \\
&+\mathbb{E}_{z \sim p_{\theta}(z|x)}[-\log{q_{\phi}(y|z)}],
\end{aligned}
\end{equation}
where ${q_{\phi}(y|z)}$ means a variational approximation of $p(y|z)$, $p_{\theta}(z|x)$ means an estimate of the posterior distribution of $z$, and $r(z)$ means an estimate of the prior distribution of $z$. We consider $p_{\theta}(z|x)$ and $r(z)$ as a normal distribution, especially $r(z)$ as standard normal distribution for computation of analytic Kullback-Leibler (KL) divergence. We parametrize those distributions with the neural network, then $p_{\theta}(z|x)$ and $q_{\phi}(y|z)$ can be interpreted as an encoder and decoder that shares compressed latent representation $z$. Since the second term of the VIB objective function represents prediction loss, we implement the objective function in the form of adding cross-entropy loss to the KL divergence between the estimate of posterior and prior.

\subsection{Wav2vec 2.0 + VIB architecture}
Figure \ref{fig:wav2+VIB} shows the overall architecture of our proposed anti-spoofing system. Wav2vec 2.0 model is pretrained with the self-supervised scheme, and suppose that the model is parametrized with parameters $\psi$. The input raw waveform $s$ is fed into the pretrained wav2vec 2.0 model $f_{\psi}(.)$, and converted into speech embedding $f_{\psi}(s)$ after pooling contextualized representations. Speech embedding $x=f_{\psi}(s)$ passes multi-layer perceptron (MLP) $g(.)$ to decrease dimension. After MLP, two linear layers yield posterior distribution mean $\mu(g(x))$ and variance $\Sigma(g(x))$, respectively. Using these statistics, we can estimate the posterior distribution of $z$ parametrized with $\theta$ as $p_{\theta}(z|x)$ since we supposed prior $r(z)$ and parametrized posterior $p_{\theta}(z|x)$ as a normal distribution. We use the reparameterization trick in \cite{kingma2013auto} by setting $z$ as $\mu(g(x))+\epsilon \odot \Sigma(g(x))$, where $\epsilon \sim \mathcal{N}(0,I)$ to take gradients. From $p_{\theta}(z|x)=\mathcal{N}(z|\mu(g(x)),\Sigma(g(x)))$, we can sample compressed speech representation $z$, and feed the representation to another MLP which is the classifier that refers to the parametrized distribution of $q_{\phi}(y|z)$ predicting the logits that decide whether the speech $s$ is spoofed or not. Note that, representation $z$ contains noise since it is sampled from the posterior distribution with variance $\Sigma(g(x))$. This stochasticity forces the network not to learn superficial information which depends on the dataset but generalized information for anti-spoofing task.

\section{Experimental setup}

\subsection{Database and metrics}

We experiment based on ASVspoof 2019 database \cite{wang2020asvspoof} logical access (LA) partition, which corresponds to a synthetic spoofing attack scenario. LA partition is divided into three subsets: train, development, and evaluation. Each subset consists of genuine speech and spoofed speech. Spoofed speech in train and development subsets is generated from 6 different TTS/VC algorithms, and spoofed speech in the evaluation subset is generated from 13 different TTS/VC algorithms.

Additionally, we use ASVspoof 2015 database \cite{wu15e_interspeech} and ASVspoof 2021 database \cite{yamagishi21_asvspoof} LA partition for cross-dataset setting. Since the model trained with ASVspoof 2019 database is evaluated with another database in the cross-dataset setting, only the evaluation subset of each database is used. Each database contains spoofed speech generated from various VC and TTS algorithms along with genuine speech. ASVspoof 2021 LA evaluation subset includes spoofed speech generated from the same algorithms with ASVspoof 2019 LA evaluation subset, but contains degraded speech with encoding or transmission artifacts from the real telephony system.

We evaluate our proposed model with two metrics: minimum normalized tandem detection cost function (min t-DCF) \cite{kinnunen18b_odyssey} and equal error rate (EER).

\subsection{Baseline systems}
We implement two baseline systems: RawNet2 system \cite{9414234} and AASIST system \cite{Jung2021AASIST}, and compare against our proposed system in low-resource and cross-dataset settings. 

RawNet2 is an end-to-end anti-spoofing system that is selected as the baseline system in ASVspoof 2021 challenge. The RawNet2 system consists of residual blocks with gated recurrent unit (GRU) layers after convolution layer with sinc filters. We reproduce the RawNet2 baseline system based on the implementation details in \cite{9414234} and ASVspoof 2021 official GitHub repository\footnote{https://github.com/asvspoof-challenge/2021}.

AASIST is an anti-spoofing system that has an architecture of integrated spectral and temporal graph attention network. The AASIST system shows state-of-the-art performance in \cite{Jung2021AASIST} compared to other best performing anti-spoofing systems. We implement the AASIST baseline system based on the detailed implementations in \cite{Jung2021AASIST} and official PyTorch implementation in the GitHub repository\footnote{https://github.com/clovaai/aasist}.

\subsection{Implementation details}
All implementations and experiments are conducted with PyTorch framework, especially training and evaluation pipelines are constructed based on PyTorch Lightning library. We use wav2vec 2.0 base model which consists of 7 convolutional blocks and 12 Transformer blocks. Since Huggingface\footnote{https://huggingface.co/} \cite{wolf-etal-2020-transformers} provides a pretrained model for the wav2vec 2.0 base model, we adopt a wav2vec 2.0 base model from Huggingface library which is pretrained with 960-hours Librispeech \cite{panayotov2015librispeech} database. Output sequences of wav2vec 2.0 are mean-pooled along with the time-axis to make 768 units of speech embedding.

MLP module in VIB architecture consists of three linear layers with 768, 640, and 512 hidden units with ReLU non-linear activations. After the MLP module, each linear layer for $\mu(g(x))$ and $\Sigma(g(x))$ has 256 hidden units and the compressed representation $z$ has 256 dimensions. We also use two linear layers to approximate distribution $q_{\phi}(y|z)$. Similar to Bowman \textit{et al}. \cite{bowman2016generating}, we use annealing strategy for $\beta$, changing it as min(1, epoch $\times$ 0.0001) during training. We average 5 posterior samples from $p_{\theta}(z|x)$ in training and use only mean samples in evaluation.

Overall architecture is trained with the train subset of ASVspoof 2019 LA database. We use weighted cross-entropy loss since the amount of each speech in the train subset is unbalanced. We assign weights of 0.9 and 0.1 to genuine and spoofed speech, respectively. For batch training, custom data collator that dynamically pads the inputs with the length of the longest input in the batch is used with a mini-batch size of 8. The AdamW \cite{loshchilov2018decoupled} optimizer is used with an initial learning rate $10^{-6}$ and the learning rate is scheduled by reducing the factor of 0.1 when the EER of the development subset does not decrease during 8 epochs. The maximum number of training epochs is 100.

\section{Results}

\subsection{Comparison with recent anti-spoofing systems}
In Table \ref{table_compare_SOTA}, we compare the proposed systems with other recent state-of-the-art anti-spoofing systems in terms of min t-DCF and EER. All systems in the table, including the proposed systems, are trained with ASVspoof 2019 LA train subset and evaluated on ASVspoof 2019 LA evaluation subset. We report min t-DCF and EER of our best-performing proposed systems. We can see that using the wav2vec 2.0 pretrained model exceeds the previous anti-spoofing systems by simply adding a classifier consisting of two linear layers. In addition, adding the VIB module between the linear classifier and the wav2vec 2.0 model improves performance about 28\% and 15\% in terms of min t-DCF and EER, respectively. This shows that the VIB effectively suppresses redundant information in the feature of the pretrained model, leading to improvement of generalization ability.

\begin{table}[hbt!]
  \caption{Performance comparison with state-of-the-art systems in terms of min t-DCF and EER (\%). All systems are trained and evaluated with the ASVspoof 2019 database LA partition.}
  \label{table_compare_SOTA}
  \centering
  {\footnotesize
  \begin{tabular*}{\columnwidth}{c|c|cc}
    \toprule
    System & Architecture & min t-DCF & EER (\%)            \\
    \midrule
    Luo \textit{et al}. \cite{9414670} & Capsule Network & 0.0538 & 1.97\\
    Wang \textit{et al}. \cite{wang21fa_interspeech} & LCNN-LSTM & 0.0524 & 1.92\\
    Chen \textit{et al}. \cite{chen20_odyssey} & ResNet18 & 0.0520 & 1.81\\
    Ge \textit{et al}. \cite{ge21_asvspoof} & Raw PC-DARTS & 0.0517 & 1.77\\
    Hua \textit{et al}. \cite{hua2021towards} & Res-TSSDNet & 0.0481 & 1.64 \\
    Zhang \textit{et al}. \cite{zhang21da_interspeech} & SENet & 0.0368 & 1.14 \\
    Tak \textit{et al}. \cite{tak21_asvspoof} & RawGAT-ST & 0.0335 & 1.06 \\
    Jung \textit{et al}. \cite{Jung2021AASIST} & AASIST-L & 0.0309 & 0.99 \\
    Jung \textit{et al}. \cite{Jung2021AASIST} & AASIST & 0.0275 & 0.83 \\
    Proposed & Wav2vec 2.0 & 0.0149 & 0.47 \\
    Proposed & Wav2vec 2.0 + VIB & \textbf{0.0107} & \textbf{0.40} \\
    \bottomrule
  \end{tabular*}
  }
\end{table}

\begin{table*}[hbt!]
    \centering
    \caption{Results of baseline systems and proposed systems in low-resource setting in terms of EER (\%). All systems are trained with the corresponding percentage of training data from the ASVspoof 2019 database LA partition. We report the performances by averaging EERs from three experiments with different random seeds.}
    \label{table_lowresource}
    \begin{tabular}{c|c|ccccccc}
        \toprule
        \multirow{2}{*}{System} & \multirow{2}{*}{Architecture} & \multicolumn{7}{c}{Percentage of training data}                                                               \\
                    &         & 1\%   & 3\%   & 5\%   & 10\%  & 25\% & 50\% & 100\% \\
        \midrule            
        Tak \textit{et al}. \cite{9414234}  & RawNet2 & 21.71 & 17.95 & 15.77 & 11.46 & 9.78 & 6.00 & 5.46  \\
        Jung \textit{et al}. \cite{Jung2021AASIST} & AASIST  & 16.90 & 6.16  & 5.32  & 4.25  & 2.98 & 1.65 & 1.26  \\
        Proposed                & Wav2vec 2.0                   & 3.59          & 2.01          & \textbf{0.89} & 0.78          & \textbf{0.51} & 0.56          & 0.50          \\
        Proposed                & Wav2vec 2.0 + VIB             & \textbf{3.39} & \textbf{1.67} & 0.95          & \textbf{0.66} & 0.56          & \textbf{0.46} & \textbf{0.42} \\
        \bottomrule
    \end{tabular}
\end{table*}

\subsection{Evaluation in low-resource setting}
In Table \ref{table_lowresource}, we report the performances of baseline systems and proposed systems in a low-resource setting. Each column labeled as a percentage means the EER results of the trained system, evaluated with ASVspoof 2019 LA evaluation subset. All systems are trained with the corresponding percentage of ASVspoof 2019 LA train subset. We implement a low-resource setting by sampling utterances randomly in the training data with a certain percentage according to each speaker and spoofing type to keep the data distribution constant. Overall, all systems show deteriorating performance as the amount of training data decreases. In particular, in the case of baseline systems, the EER nearly doubles when the amount of data reaches 25\% and deterioration becomes severe as the amount of data reduces. On the other hand, systems based on wav2vec 2.0 show similar performance to full data performance with only using 25\% training data and show higher performances than the best performance of other baseline systems with only using 5\% data. In addition, the performance improves with using VIB in all cases except 5\% and 25\%. Results show that transferring knowledge from wav2vec 2.0 pretrained model to an anti-spoofing system effectively addresses low-resource scenarios and the VIB strengthens this robustness in terms of sizes of training data by suppressing irrelevant information.

\subsection{Evaluation in cross-dataset setting}

\begin{figure}[hbt!]
  \centering
  \begin{subfigure}[b]{0.49\linewidth}
    \includegraphics[width=\linewidth]{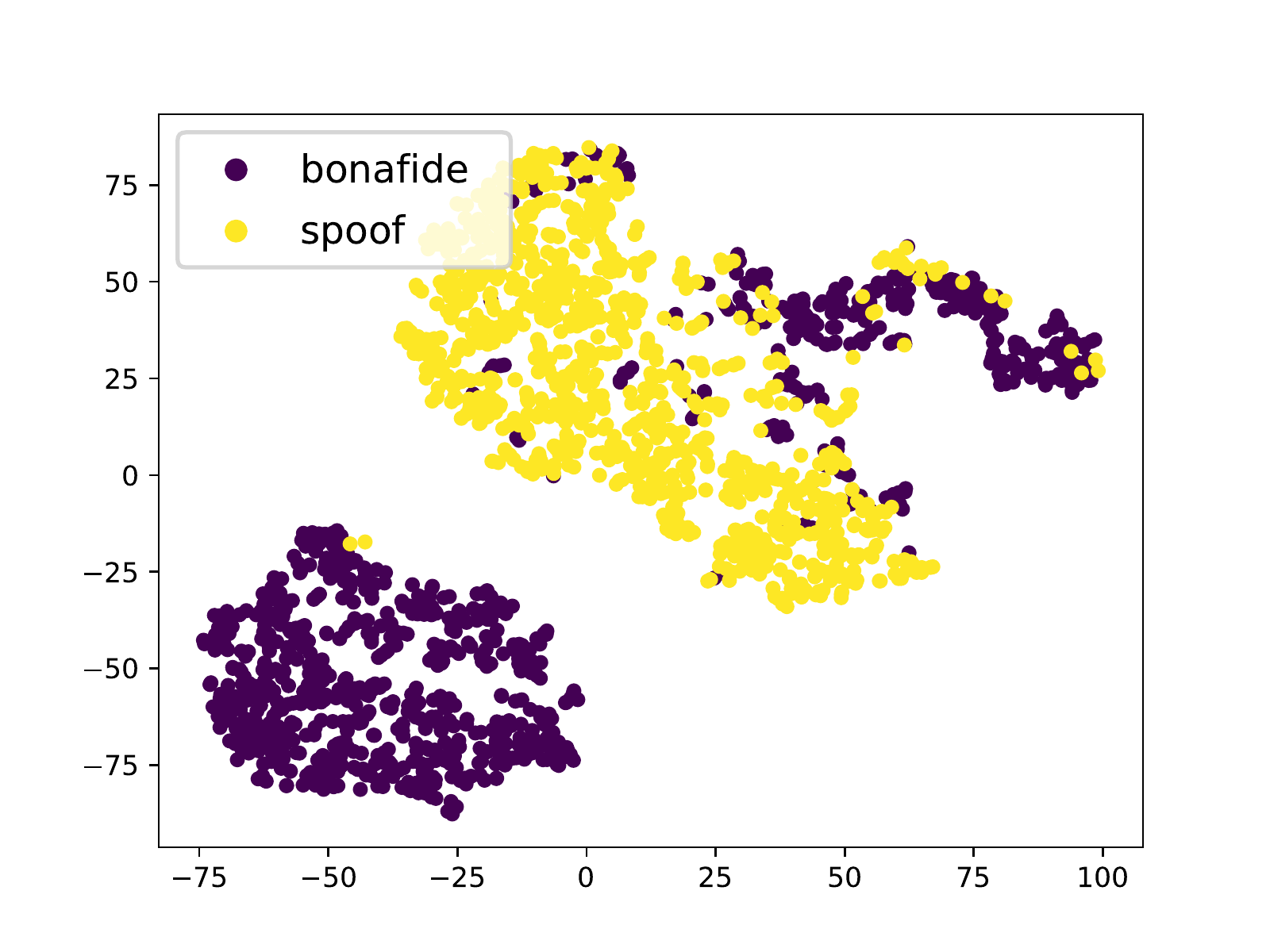}
    \caption{Wav2vec 2.0}
  \end{subfigure}
  \begin{subfigure}[b]{0.49\linewidth}
    \includegraphics[width=\linewidth]{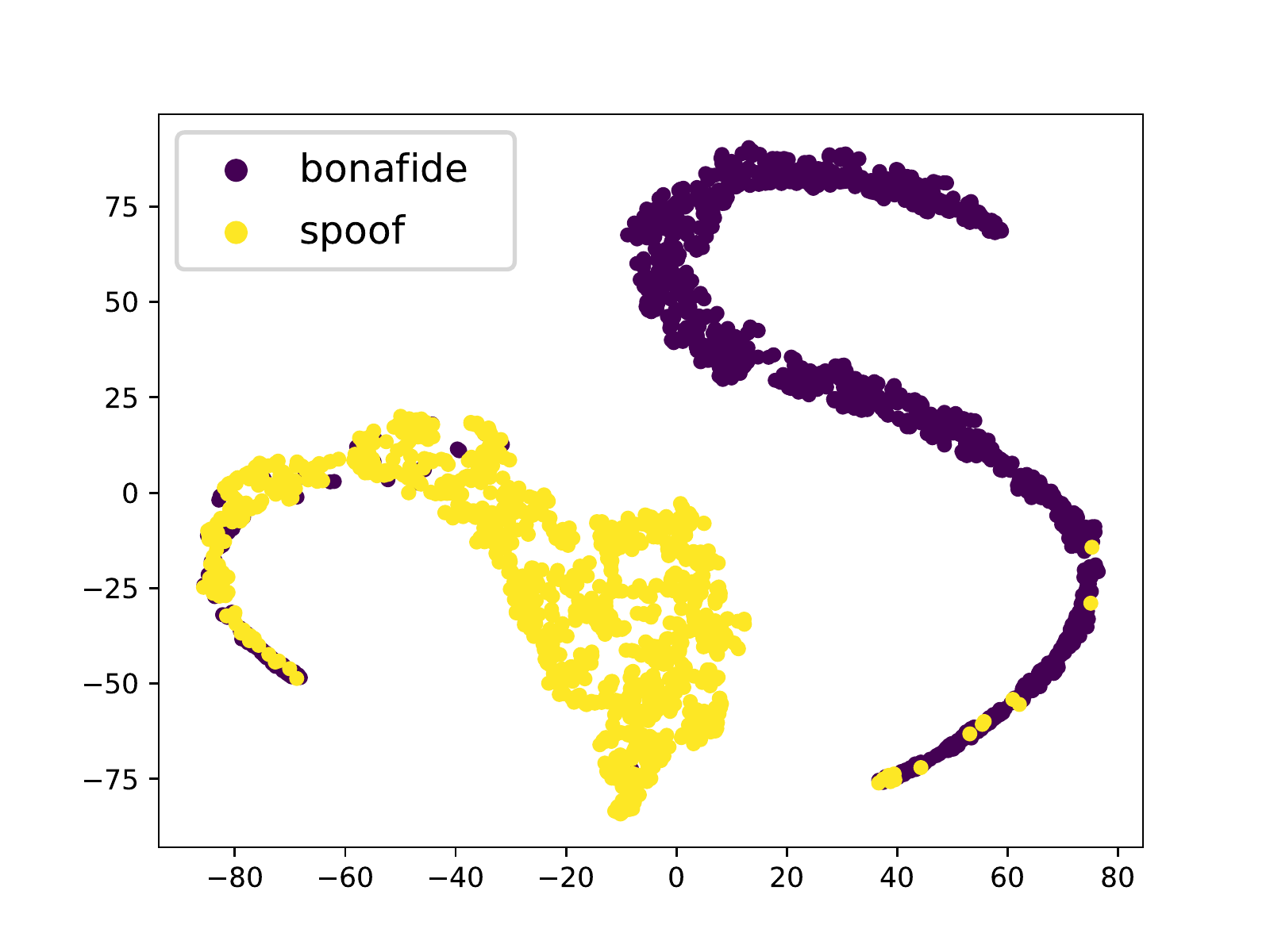}
    \caption{Wav2vec 2.0 + VIB}
  \end{subfigure}
  \caption{Visualization of t-SNE embeddings from proposed systems in cross-dataset setting evaluated on the ASVspoof 2021 LA evaluation subset. Each color represents whether the speech is genuine (bonafide) or spoofed.}
  \label{fig:t-SNE}
\end{figure}

In Table \ref{table_cross_dataset}, we compare the performances of baseline systems and proposed systems in a cross-dataset setting. All systems are trained with ASVspoof 2019 LA train subset and evaluated with the evaluation subset of another database. Numbers in each column labeled as database mean the EER results of the systems evaluated with the evaluation subset of the corresponding database. Overall results show that the performance is degraded in cross-dataset setting compared to the performance evaluated with ASVspoof 2019 LA evaluation subset. Particularly, in the case of ASVspoof 2021 LA, although it is composed of the same spoofing attacks as in ASVspoof 2019 LA, the performance is deteriorated due to the degraded speech from transmission artifacts. We can see that the system using wav2vec 2.0 outperforms baseline systems in all cross-dataset settings. Furthermore, using VIB also improves performance in all cross-dataset settings, especially performance is improved by about 44\% in the case of ASVspoof 2021 LA. The improvements of our proposed system support that our method improves generalization performance on the degraded data or data from different distributions by learning general information rather than irrelevant and superficial information from the training data. 

Also, in Figure \ref{fig:t-SNE}, we visualize embeddings from each proposed system evaluated on ASVspoof 2021 LA evaluation subset by applying t-SNE \cite{van2008visualizing}. We can see that embeddings of the wav2vec 2.0 + VIB system are more discriminative and form a well-constructed manifold in embedding space than the wav2vec 2.0 system. This result shows that our proposed VIB system learns general representation for discriminating genuine and spoofed speech, and shows the reason for robustness against various unseen spoofing attacks and environments.  

\begin{table}[h!]
    \centering
    \caption{Results of baseline systems and proposed systems in cross-dataset setting in terms of EER (\%). All systems are trained with the ASVspoof 2019 LA database, and evaluated with the ASVspoof 2015 and ASVspoof 2021 LA databases.}
    \label{table_cross_dataset}
    {\footnotesize
    \begin{tabular*}{\columnwidth}{c|cc}
        \toprule
        System   & ASVspoof 2015 & ASVspoof 2021 LA \\
        \midrule
        RawNet2 \cite{9414234}           & 6.62                   & 9.50                      \\
        AASIST \cite{Jung2021AASIST}            & 3.22                   & 10.90                     \\
        Wav2vec 2.0       & 2.02                   & 8.81                      \\
        Wav2vec 2.0 + VIB & \textbf{1.52}          & \textbf{4.92}             \\
        \bottomrule
    \end{tabular*}
    }
\end{table}

\section{Conclusion}

We proposed a transfer learning scheme based on the wav2vec 2.0 pretrained model with variational information bottleneck (VIB) for the anti-spoofing task. Incorporating VIB to wav2vec 2.0 model helps to find generalized representation which is relevant to the task, suppressing irrelevant and superficial information. Evaluation result on the ASVspoof 2019 LA database showed state-of-the-art performance with only using a simple classifier. In addition, we showed that our system improves generalization performance in low-resource and cross-dataset settings as well. In the future, we will investigate the effectiveness of the proposed method using other recent pretrained models and sophisticated classifiers.

\section{Acknowledgements}

This work was supported by Institute of Information \& communications Technology Planning \& Evaluation(IITP) grant funded by the Korea government(MSIT) (No.2021-0-00575, Development of Voicepishing Prevention Technology Based on Speech and Text Deep Learning)
\bibliographystyle{IEEEtran}

\bibliography{mybib}

\end{document}